\documentstyle[aps,twocolumn,epsf,psfig]{revtex}
\begin{document}
\draft

\title{Is the direct observation of electronic coherence in electron 
transfer reactions possible?}

\author{Andreas Lucke,${}^{1,2}$ C.H. Mak,${}^1$ Reinhold Egger,${}^2$
Joachim Ankerhold,${}^{2,3}$ Juergen Stockburger,${}^1$ and 
Hermann Grabert${}^2$}

\address{
${}^1$Department of Chemistry, University of Southern California,
Los Angeles, CA 90089-0482, USA\\
${}^2$Fakult\"at f\"ur Physik, Albert-Ludwigs-Universit\"at, 
Hermann-Herder-Stra{\ss}e 3, D-79104 Freiburg, Germany \\
${}^3$Department of Chemistry, Columbia University,
New York, NY 10027, USA}

\date{Date: \today}
\maketitle

\begin{abstract}
The observability of electronic coherence in electron 
transfer reactions is discussed. 
We show that under appropriate circumstances
large-amplitude oscillations can be found
in the electronic occupation probabilities. The initial
preparation of the system is of crucial importance for
this effect, and we discuss conditions under which experiments
detecting electronic coherence should be feasible. 
The Feynman-Vernon influence functional formalism is extended
to examine more general and experimentally relevant
initial preparations. 
Analytical expressions and path integral quantum dynamics simulations
were developed 
to study the effects of various initial preparations 
on the observability of electronic coherence.
\end{abstract}

\narrowtext

\section{Introduction}

The role of quantum coherence in the dynamics of molecular systems in 
the condensed phase is still largely unclear.  According to conventional 
wisdom, 
quantum coherence should have little influence on condensed phase dynamics
because the dephasing time of any coherent process is expected to be very
short for systems with any appreciable number of degrees of freedom.
The extent to which conventional wisdom is correct depends on many 
parameters such as the effective mass of the degree of freedom under
consideration, its coupling strength to the bath, and the timescale of the
bath motions.  Therefore, the understanding of quantum coherence in 
condensed phase molecular systems can be a rather complicated issue.

The experiments of Vos et al.\cite{vos} on the photosynthetic reaction 
center were one of the first in which vibrational coherence was directly
observed in an electron transfer (ET) system.  Similar 
experimental observations in other systems have been reported
since 
\cite{stanley-boxer,ar-voe-sche,koch-math,chach-sund,wynne-reid-hoch,reid}.
The most surprising aspect of the discovery is that quantum 
coherence is directly detectable and is actually preserved on a rather long 
timescale even in condensed phase systems with ultrafast solvent 
dynamics.  These direct observations of vibrational coherence in ET 
systems have led to speculations concerning the possible influence 
coherent quantum dynamics may have on the ET dynamics itself.  
There have been a number of theoretical studies on this 
subject \cite{jean-friesner,alden,ben-lev-fle,coalson,ev-coal}.
However, experiments have so far been unable to establish a clear 
connection between vibrational coherence and ET rates.  

In this paper, we will focus on electronic coherence in ET reactions.
Electronic coherence, if present, would have a more profound effect on the
ET dynamics than vibrational coherence.  
In the absence of the environment, an ET system when put into 
a state represented by a superposition of two or more electronic 
eigenstates will evolve coherently.  
This simple picture of electronic coherence becomes immensely muddled 
when the environment is taken into account, because the coherent 
electronic motions can be dephased by the nuclear motions in the environment.  
In fact, in most situations the electronic motions
are so strongly coupled to the nuclear motions that the combined system is 
better described by the vibronic states.  In this way the electronic motions
associated with the electronic coherence will acquire a larger effective
mass, and the crucial experimental question is whether electronic 
coherence can be 
observed directly in ultrafast experiments on a timescale fast enough to 
beat the dephasing time.  A recent experimental attempt by Reid et 
al.\cite{reid} to detect electronic coherence in biruthenium mixed-valence 
compounds has failed to yield any direct evidence because the dephasing 
timescale seems to be too short even for the 20~fs time resolution of the 
experiment.  Given this negative result, the possibility of a direct 
experimental detection of electronic coherence may appear unlikely.

The purpose of this paper is to re-examine the theoretical issues surrounding
the observability of electronic coherence in ET systems.
Contrary to prevailing skepticism, we believe 
that there is no intrinsic experimental limitation that fundamentally
prevents electronic coherence from being detected. We employ simple arguments
to show that the major difficulties encountered in the direct experimental 
observation of electronic coherence are related to how the system is prepared 
initially in the experiment.  By using various methods of nonequilibrium 
initial preparation, the duration of time within which electronic coherence 
is preserved as well as the amplitude of the coherent oscillations
can be increased dramatically.  
The validity of these simple ideas have been verified by dynamical
path integral simulations.
Our analyses suggest that electronic coherence can be most
easily observed in adiabatic ET systems in which the electronic coupling
matrix element is large compared to the frequency of typical solvent motions.
ET systems that satisfy these requirements are often found in 
mixed-valence compounds \cite{creutz}, which have typical electronic coupling
matrix elements of the order of several hundred wavenumbers.

The organization of the paper is as follows.
Section \ref{secii} presents a simple analysis of electronic 
coherence in an ET experiment.  Using this analysis, we study the effects
of various preparation methods on the strength and the dephasing
time of the electronic coherence in the adiabatic limit.
Based on these results, we propose a number of specific ways to engineer the
initial preparation of the system so that electronic coherence could be 
more easily detected.
Section \ref{seciii} develops these qualitative ideas further by a detailed 
quantitative analysis using the two-state spin-boson Hamiltonian as a
model for mixed-valence compounds. 
The  Feynman-Vernon influence functional method is extended to
allow for more general
initial preparations, and the concept of an ``equivalent external
field'' is developed.
Accurate quantum dynamics simulations show that the 
qualitative predictions of the simple theories from Section \ref{secii} are 
generally preserved when the full solvent dynamics is taken into account.
For completeness, we also include in Section \ref{seciv} a 
brief summary of results for the nonadiabatic region.

\section{Electronic Coherence for Adiabatic Electron Transfer}
\label{secii}

\subsection{Standard Preparation}
\label{sec:tp}
In the absence of the solvent, if the two electronic states before and after
the charge transfer are denoted $\vert L \rangle$ and $\vert R \rangle$,
the electronic Hamiltonian can be written in the tight-binding
approximation as 
\begin{eqnarray}
H_e = &-& \frac{\hbar\Delta}{2} \left( \vert L \rangle \langle R \vert +
                  \vert R \rangle \langle L \vert \right) \nonumber \\
                &+& \frac{\hbar\epsilon}{2}  \left(
                  \vert L \rangle \langle L \vert -
                  \vert R \rangle \langle R \vert \right)\;,
\end{eqnarray}
where $\epsilon$ is a bias related to the intrinsic difference in the
redox free energies $\Delta G^0$ of the two electronic states.
In condensed phase ET systems, this electronic system is immersed in a
bath.  In most bimetallic mixed-valence compounds, the electronic coupling 
matrix elements $\hbar\Delta/2$ are large enough for them to be classified 
as adiabatic ET systems\cite{creutz}.
This means that the timescale of the pure electronic motions,
set by the electronic matrix element $\hbar\Delta/2$, is much shorter than the
timescale of the bath motions.  Although the frequencies of bath motions 
often cover a broad and almost continuous spectrum, there
is often a typical bath frequency $\omega_c$ such that motions with frequencies
significantly higher than $\omega_c$ are absent in the bath\cite{remark2}.  
The adiabatic regime is then defined by the condition 
$\Delta \gg \omega_c$.  

Under the condition $\Delta \gg \omega_c$,
we can invoke the Born-Oppenheimer approximation
and arrive at the adiabatic limit of Marcus theory 
\cite{marcus,chandlerbook,chandler-lfgt}.  In this
limit, the bath polarization ${\cal E}$ can be treated as a classical 
coordinate, which couples to the dipole moment of the electronic 
state.  The electronic Hamiltonian is modified by this coupling and 
it becomes 
\begin{eqnarray}
H_e({\cal E}) = &-& \frac{\hbar\Delta}{2} \left( \vert L \rangle 
\langle R \vert + \vert R \rangle \langle L \vert \right) \nonumber \\
          &+& \left(\frac{\hbar\epsilon}{2} + \mu {\cal E}\right)  \left(
  \vert L \rangle \langle L \vert - \vert R \rangle \langle R \vert \right)\;,
\label{eq:HofE}           
\end{eqnarray}
where $\mu$ is one-half the difference in the dipole moments of the two 
electronic states.
In the adiabatic limit, the bath polarization is slow, so that the
Born-Oppenheimer approximation gives as a function of the bath polarization
two electronic eigenstates with energies
\begin{equation} \label{eigenstates}
E_{\pm}({\cal E}) = \pm \sqrt{\left( \frac{\hbar\Delta}{2} \right)^2 + 
\left (\frac{\hbar\epsilon}{2} + \mu {\cal E} \right)^2}\;.
\end{equation}
Figure~\ref{fig1}(a) 
shows the energies of the two eigenstates as a function of the
bath polarization ${\cal E}$ for a symmetric system ($\epsilon =0$).

Within Marcus theory the bath polarization obeys linear response.  Therefore,
the orientation of the bath polarization when the bath is uncoupled to the
electronic system has an intrinsic gaussian distribution
\begin{equation} \label{rho0}
\rho_0({\cal E}) = Z_{0}^{-1} \exp \left( -\beta \mu^{2}{\cal E}^2/
\lambda \right)\;,
\end{equation}
where $\beta = 1/k_B T$, $Z_{0}=\int d{\cal E} \exp(-\beta \mu^2 
{\cal E}^2/\lambda)= \sqrt{\pi\lambda/\mu^2\beta}$,  and 
$\lambda$ is the classical reorganization energy.

In a normal ET experiment, we imagine an initial preparation in which
we hold the electron in state $\vert L \rangle$ and thermalize the 
bath polarization to it.  
In other words, the bath is prepared initially with a distribution 
\begin{equation} \label{eq:rhoE}
\rho_{\rm S}({\cal E}) = Z_{\rm S}^{-1}
\exp \left[ -\beta \left( \mu^{2}{\cal E}^2/\lambda + \mu {\cal E}
\right) \right]\;.
\end{equation}
Related to the intrinsic bath polarization distribution (\ref{rho0}), 
this distribution has the same gaussian shape, but its maximum is shifted 
to $\mu {\cal E}=-\lambda/2$.  This is the usual initial preparation one 
considers in theoretical descriptions of ET.
Here, we shall always refer to it as the ``standard preparation''.

At time $t=0$, we release the electron and observe its dynamics.
Because the bath polarization is slow compared to the electron, we can 
compute the early-time dynamics of the electron assuming that the 
bath is stationary.  This corresponds to solving for the dynamics of
the electron using the two eigenstates (\ref{eigenstates})
for every position of the bath polarization ${\cal E}$, and then averaging
the result over the distribution function $\rho_{\rm S}({\cal E})$ 
in Eq.~(\ref{eq:rhoE}).
This procedure is illustrated schematically in Fig.~\ref{fig1}(a).
 Due to the gaussian
preparation $\rho_{\rm S}$, only the polarizations between the two dashed 
lines
are significantly populated.  For each ${\cal E}$, the dynamics of a biased
but free two-level system can be solved giving $P_L(t;{\cal E})$, 
the transient population on the $L$-state, with the initial condition set
to $P_L(0;{\cal E}) = 1$.
An average over ${\cal E}$ finally produces the observed $P_L(t)$:
\begin{equation} \label{plt}
P_L(t) = \int_{-\infty}^{\infty} d{\cal E} \rho_{\rm S}({\cal E}) 
P_L(t;{\cal E})\;.
\end{equation}
The solid line in Fig.~\ref{fig2} illustrates a typical result of 
Eq.~(\ref{plt}) for a symmetric system, $\epsilon = 0$.
(In Fig.~\ref{fig2}, time is measured in units of an inverse frequency, 
$1/\omega_{c}$, 
which for now has no special meaning other than setting a timescale.)
As expected for the case of a standard preparation, the 
$L$-state population remains approximately unity, indicating a very slow 
ET rate from the $L$- to the $R$-state.  In Section~\ref{sec:qmc}, 
we will verify that the qualitative behavior predicted by
this adiabatic calculation is indeed correct using a series of exact
quantum simulations.

\subsection{Nonstandard Preparations}
\label{sec:ntp}

In this section, we will examine several experiments in which the initial 
preparation of the bath is carried out by a distribution function different 
from $\rho_{\rm S}$.  We call these ``nonstandard preparations''.
In these experiments, the electron is held in the $L$-state for time $t<0$
as in the standard experiment, but the bath is prepared by a 
generalized distribution function
\begin{equation} \label{eq:rhoNE}
\rho_{\rm NS}({\cal E};\bar{q},\bar{\beta}) =  Z^{-1}_{\rm NS}
\exp \left[ -\bar{\beta} \left( \mu^{2}{\cal E}^2/\lambda - \bar{q} {\cal 
E}\mu \right) \right],
\end{equation}
in which the bath polarization ${\cal E}$ is equilibrated to an effective 
electronic dipole moment of magnitude $\bar{q}\mu$ instead of $\bar{q}=-1$
as in $\rho_{\rm S}$. In addition, 
the initial bath polarization distribution can
have a different width, i.e. be at an effective temperature $\bar{\beta}$.

Within Marcus theory, 
the initial bath preparation specified by $\rho_{\rm NS}$ represents the 
proper description for systems which are not necessarily thermalized with 
the bath coupled to the electron held fixed in the $L$-state.
The distribution in Eq.~(\ref{eq:rhoNE}) 
differs from that in Eq.~(\ref{eq:rhoE}) by a horizontal shift 
in $\mu {\cal E}$ by an amount of
$(\bar{q}+1)\lambda/2$. If $\bar{\beta} \neq \beta$, the distributions 
also have a different width. The following discussion focuses 
predominantly on the effect of the position of the distribution only, 
not its shape. Therefore, 
we will concentrate on those nonstandard distributions with 
$\bar{\beta}=\beta$ and refer to them specifically as ``shifted bath
preparations". 
In the following, we suggest several possible scenarios which give rise 
to an initial preparation described by Eq.~(\ref{eq:rhoNE}).
Similar nonstandard initial preparations have been considered in 
Refs.\cite{coalson,ev-coal} for the nonadiabatic regime, but to our knowledge
nonstandard initial preparations have not been studied for the adiabatic regime
before.

\begin{figure}
\setlength{\unitlength}{1cm}
\begin{center}
\leavevmode
\epsfxsize 8cm 
\epsffile{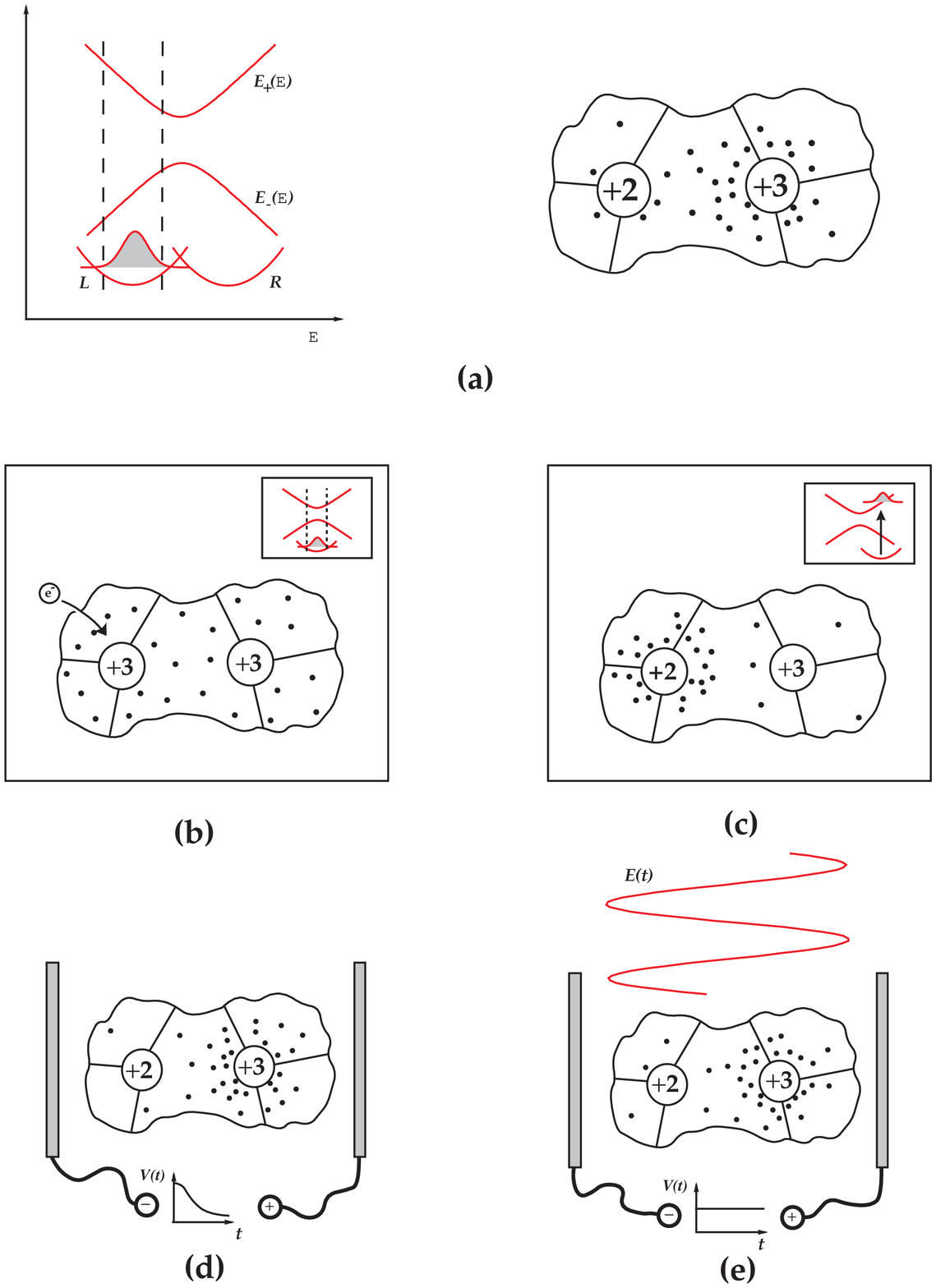}
\end{center}
\caption[]{\label{fig1}
(a) The electronic energy eigenvalues $E_{\pm}({\cal E})$ as a 
function of the bath polarization ${\cal E}$. 
The shaded gaussian represents 
$\rho_{\rm S}({\cal E})$ that arises from a parabolic free energy surface 
(Marcus parabola). The mixed-valence compound for this standard 
preparation is sketched. The excess electron sits on the left metal 
center, and the bath polarization has adjusted to it (indicated by 
the density of dots).
(b) Experimental situation where we expect 
large electronic coherence effects.
For all times $t<0$, there is no electric 
dipole moment and therefore $\langle{\cal E} \rangle = 0$, 
depicted by the uniform distribution of dots. At time $t=0$ an electron is 
injected into the left atom putting the electron in 
state $|L\rangle$, but with $\rho_{\rm NS}({\cal E}; \bar{q}=0, \beta)$ 
(see inset).
(c) The initial preparation for backward ET in which the bath is 
equilibrated with respect to the excess electron in the acceptor.
(d) While the actual system is prepared in the standard way, the 
initial bath polarization can be compensated by a 
suitable equivalent external field, so that the ET behaves just as in 
the case of Fig.~\ref{fig1}(b).
(e) In the actual experiment, it will be necessary to 
emulate the bath preparation using an optically produced 
equivalent external field atop a static background.}
\end{figure}

\begin{figure}
\setlength{\unitlength}{1cm}
\begin{center}
\leavevmode
\epsfxsize 7cm 
\epsffile{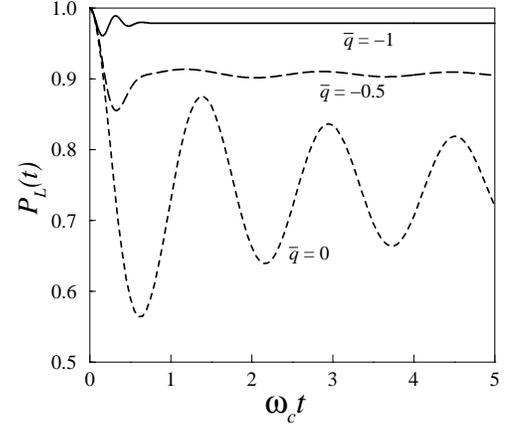}
\end{center}
\caption[]{\label{fig2} 
$P_{L}(t)$ evaluated in the adiabatic limit, for standard 
bath preparation $\rho_{\rm S}({\cal E})$ (solid curve) and shifted 
bath preparations
$\rho_{\rm NS}({\cal E};\bar{q} =-0.5,\bar{\beta}=\beta)$ (long-dashed curve)
 and $\rho_{\rm NS}({\cal E};\bar{q}=0,
\bar{\beta}=\beta)$ (short-dashed curve). The parameters used were
$\Delta=4\omega_{c}$, $\lambda = 20 \hbar \omega_{c}$ and $\hbar \beta 
\omega_{c} =3$ for a symmetric system, $\epsilon=0$.} 
\end{figure}

First, consider the experiment depicted in Fig.~\ref{fig1}(b).  Using a 
symmetric Fe(III)-Fe(III) compound as an example, we imagine an experiment
in which an extra electron is photoinjected into the ion on the left at time
$t=0$.  Prior to the injection, the electronic state has a zero dipole
moment, so the bath is initially prepared with a distribution
function $\rho_{\rm NS}$ characterized by $\bar{q}=0$ and 
$\bar{\beta}=\beta$.
Using the adiabatic approximation described in Sect.~\ref{sec:tp},
 we can now compute
the transient electronic population on the left ion $P_L(t)$ using 
$\rho_{\rm NS}$ for the bath preparation. 
 The result is shown in Fig.~\ref{fig2} 
as the short-dashed line.  Remarkably, $P_L(t)$ shows large-amplitude coherent 
oscillations.  The origin of this coherent behavior is purely electronic,
and it can be explained by referring to the inset
in Fig.~\ref{fig1}(b).
The relevant range of ${\cal E}$, depicted by the region between the
dashed lines, is now shifted to near ${\cal E}=0$.  Inside this range,
the two adiabatic electronic eigenstates are close in energy (approximately
$\Delta$ apart).  Consequently, when forced into the $L$-state which is
a superposition of the energy eigenstates, the electronic motion will exhibit a 
coherent oscillation with a frequency that is of the order of $\Delta$.

At this point, it is necessary to examine why preparing the bath thermally at 
$\bar{q}=-1$ does not seem to yield similar coherent oscillations, whereas
the nonstandard preparation does.  
The absence of oscillations in the first case is not due to  
dephasing, because so far the bath has been assumed to be static and hence 
nondissipative. There are no oscillations, simply because for 
polarizations ${\cal E}$ relevant for this initial preparation 
(provided one has similar parameters
as in Fig.~\ref{fig2}), each of the electronic states 
$|L\rangle$ and $|R\rangle$ is close to one of the two energy eigenstates,
$|E_{-}\rangle$ and $|E_{+}\rangle$, respectively, which themselves are of 
course stationary.
On the other hand, if ${\cal E} = 0$, both $|L\rangle$ and $|R\rangle$ 
are superpositions of both energy eigenstates to the same extent. Therefore,
when the initial preparation provides polarizations close to zero, we can 
expect large oscillations in the time-dependent electronic occupation
probabilities.

Though elementary, this observation is the conceptual foundation of 
the present work. We see that in a usual experiment (i.e. with the
standard preparation), lack 
of electronic oscillations is expected for reasons that 
have nothing to do with dephasing. 
The conventional wisdom that absence of electronic oscillations is 
due to rapid dephasing is not necessarily correct, because coherent 
oscillations can be suppressed by the way the system is prepared 
instead of by dephasing.
Moreover, the standard way ET experiments are carried out are not the proper
experiments for observing electronic coherence. 
We want to re-examine this issue in detail.
From our analysis of the adiabatic limit we
can understand the physical picture and try to construct experiments which 
are best positioned to detect electronic coherence directly.  
In Section~\ref{seciii} we go beyond the adiabatic limit and 
formulate the problem rigorously 
including the quantum dynamics of the bath as well 
as shifts in the initial preparations. With that formulation,  
we will verify that even for moderately large 
$\Delta/\omega_{c}$, the adiabatic predictions remain qualitatively 
correct. 

We can imagine a few more experimental situations in which nonstandard 
initial preparations arise. In the experiments by Vos et al. 
\cite{vos}, for example,
the donor state is populated by a fast laser pulse which leaves the 
environment essentially unchanged during excitation. Therefore, the initial 
bath polarization distribution corresponds to a gaussian wavepacket
that is equilibrated with respect to an electronic state different from
$|L \rangle$.  Such an initial preparation can be characterized by a 
$\rho_{\rm NS}$ with a $\bar{q} \neq -1$.
In Fig.~\ref{fig2}, 
we have also shown $P_{L}(t)$ for a nonstandard preparation 
with $\bar{q} = -0.5$. This could be achieved, say, by
ultrafast laser pulse excitation 
from a ground state with an electronic dipole moment 
of magnitude $-0.5\mu$.
Even for this situation, we find
that $P_{L}(t)$ (long-dashed curve 
in Fig.~\ref{fig2}) still displays some, albeit
much weaker, coherent oscillations. 

Another experiment in which a nonstandard bath preparation can be achieved
is an ultrafast back electron transfer (b-ET) experiment depicted in
Fig.~\ref{fig1}(c).  In this experiment, a fast laser pulse is used to excite a 
mixed-valence compound into the charge-transfer (CT) band to initiate a
fast b-ET reaction back to the ground state.  When the system is placed on
the CT state, the polarization
will quickly move away from the excitation region and
may not have sufficient time to equilibrate.  This situation is describable
by a nonstandard bath preparation $\rho_{\rm NS}$ with $\bar q=+1$.

The few experiments described above may not always be practically 
feasible. 
For example, in the experiment illustrated in Fig~\ref{fig1}(a), only 
in a highly idealized situation will the photoinjected electron accquire a 
pure $|L \rangle$ state instead of a superposition of $|L \rangle$ and
$|R \rangle$.
There is no doubt that the proper initial preparation 
will be tricky to establish, otherwise electronic coherence would have 
been seen already. Alternative ways to achieve similar shifted
preparations will need to be formulated.  

In Sect.~\ref{sec:eqf}, it is shown that remarkably the same
ET behavior associated with
shifted preparations of the type of $\rho_{\rm NS}$ in Eq.~(\ref{eq:rhoNE})
can also be achieved by using the standard preparation, but 
coupling the electron to a suitably chosen external 
time-dependent electric field that is tailored to mimic the effects of
the shifted bath. For example, instead of releasing the electron into a 
polarized environment, one could imagine an initially 
unpolarized bath and apply 
an external field that couples to the electric dipole moment just as the 
bath polarization would, thus producing identical ET behavior.
As the effects of the initial bath preparation dissipates, we would have to 
adjust the external field to mimic the time-dependent effects of the bath.
We call such a field the 
``equivalent external field'', because it exactly compensates for
the effects on the ET that a different initial preparation would have had.
This method is depicted schematically in Fig.~\ref{fig1}(d).
A time-dependent electric field like the one depicted in Fig.~\ref{fig1}(d)
is difficult to generate experimentally, because the decay time of 
the field must be similar to the bath relaxation time which is often
of the order of a few hundred femtoseconds. An electric field of 
sufficient strength that varies on that timescale can probably only be 
generated optically. We will show in Section~\ref{seciii}, that approximating 
the true equivalent electric field crudely by a static field plus a  
CW laser field 
and applying it to the equilibrium system can induce
electronic coherent behaviors very similar to that 
with a $\bar{q}=0$ initial preparation.
Such a field is 
schematically shown in Fig.~\ref{fig1}(e). 

Finally, there may be additional difficulties with the experimental 
observability of electronic coherence.  Even if the proper initial 
preparations or equivalent electric fields can be achieved, the coherent
oscillations in the electronic populations may be difficult to detect
experimentally.  Since the coherent oscialltaions, if present, are
expected to have a frequency $\sim \Delta$, given the best time resolution of 
currently available lasers of approximately 20~fs, this limits the systems
in which coherent ET dynamics can be observed to those having $\Delta$
smaller than a few hundred wavenumbers.

\section{Theory: Path Integral Formulation and Quantum Monte Carlo Simulations}
\label{seciii}

\subsection{Nonstandard Initial Preparations}
\label{sec:qmc}

In this section, we want to verify that the qualitative predictions from
the last section concerning the detectability of electronic coherence are
still correct when the full quantum dynamics of the bath are taken into
account.

ET with a quantum mechanical bath has thoroughly been studied
theoretically in the framework of the spin-boson model 
\cite{leggett87,chandler-lfgt,weiss}. In the spin-boson model, the electronic 
part of the Hamiltonian is described by Eq.~(\ref{eq:HofE}), but the bath
polarization $\mu {\cal E} =\sum_k c_k x_k$
is represented by a linear combination of an infinite collection 
of linearly-responding effective solvent modes $\{x_k\}$ \cite{chandler-lfgt}.  
An infinite number of solvent modes is required for a proper description 
of a truly dissipative environment. The coupling constants $c_k$ that
determine the interaction strength of the solvent modes with the
electronic states are specified by a spectral density
$J(\omega)$\cite{weiss}.

Within this model, the ET dynamics, that is the time dependent population
probabilities $P_{L}(t)$ $(P_{R}(t))$ of the left (right) state can be
determined numerically exactly using a method known as the quantum Monte
Carlo (QMC) path integral method. This technique has
been successfully employed to investigate the dynamics in many ET systems
and has been described at length in many of our previous papers, 
e.g.  Refs.\cite{QMCprb,QMCjpc,QMCcpl}, 
and reviews\cite{QMCrev-acp,QMCrev-encyc}. 
We will not repeat the details here.

We imagine an ET experiment in which the electron is created in the 
donor state prior to the start of the reaction.  
If the electron is held in the donor state for a period of time long 
enough for the bath to equilibrate to it, the initial density matrix
is given by a `factorized'' form \cite{leggett87,weiss}:
\begin{equation} \label{eq:rho0} 
\rho(0) = |L\rangle \langle L| \; \;e^{-\beta(H_{B} + \mu {\cal E})}\;,
\end{equation}
where $H_B$ is the bare bath Hamiltonian.  In typical situations, the
electron is strongly solvated in the donor state, and
the initial preparation specified by Eq.~(\ref{eq:rho0}) represents
the standard way an ET experiment is often carried out \cite{remark1}.
In the classical limit, this is equivalent to the standard preparation
we considered in Sect.~\ref{sec:tp} and depicted in Fig.~\ref{fig1}(a).

Nonstandard initial conditions result when the bath does not have sufficient
time to equilibrate to the electron created in the donor state.  An example
of an experiment in which this may occur is the one depicted
 in Fig.~\ref{fig1}(b)
and considered in Sect.~\ref{sec:ntp}.
Following the photoinjection of an electron
into the donor state in Fig.~\ref{fig1}(b), 
the bath may not have enough time to 
equilibrate to the electron if the subsequent ET rate is rapid.  This would
result in a factorized initial condition similar to 
Eq.~(\ref{eq:rho0}), except
the bath is equilibrated to a zero-dipole-moment electronic system such
that the preparation is now defined by the initial density matrix
\begin{equation} \label{eq:rho1}
\rho(0) = |L\rangle \langle L| \; \;e^{-\beta H_{B}}\;.
\end{equation}
This gives rise to the nonstandard shifted bath preparation we have 
considered in Sect.~\ref{sec:ntp}.
In general, we can represent any factorized system preparation by an 
initial density matrix in the form
\begin{equation} \label{eq:rhoq}
\rho(0;\bar q) =
|L\rangle \langle L| \; \;e^{-\beta(H_{B} - \bar q \mu {\cal E})}\;,
\end{equation}
where the parameter $\bar q$ determines whether the bath preparation is
standard or shifted. Obviously,
$\bar q=-1$ corresponds to the standard preparation
and $\bar q=0$ to the experiment of Fig.~\ref{fig1}(b).
The same $\rho(0)$ can also be used to describe the initial preparation in
the b-ET experiment in Fig.~\ref{fig1}(c),
the appropriate value of $\bar q$ there 
being $+1$. For a laser pulse excitation from the ground-state
Born-Oppenheimer surface, at least in 
principle, any value of $\bar{q}$ can be generated.

The computer programs from the previous QMC studies 
\cite{QMCprb,QMCjpc,QMCcpl,QMCrev-acp,QMCrev-encyc}
were modified to 
include the nonstandard preparation defined by Eq.~(\ref{eq:rhoq}).  A series
of QMC simulations were performed for a number of ET parameters and 
different nonstandard preparations to examine the influence of shifted
bath preparations on the time-dependent occupation probability $P_L(t)$.
In particular, we looked for conditions under which large-amplitude
oscillations were evident in $P_L(t)$.

Since the bath obeys linear response, all effects of the bath on the
electronic states can be described by the bath polarization correlation 
function, which is related to the spectral density $J(\omega)$ by
\begin{equation} \label{eq:EE}
\langle \delta {\cal E}(t) \delta {\cal E}(0) \rangle
       = \frac{\hbar}{(2 \mu)^{2}\pi} \int_{0}^{\infty} d \omega J(\omega)
         \frac{\cosh(\omega [\hbar \beta/2 - it])} {\sinh(
         \omega \hbar \beta/2)}\;.
\end{equation}
For most ET systems, a sufficiently realistic choice for $J(\omega)$ is
the ohmic spectral density with an exponential cutoff,
\begin{equation} \label{eq:J}
J(\omega) = (2 \pi \hbar \alpha) \omega e^{- \omega/\omega_{c}}\;,
\end{equation}
where the dimensionless
friction parameter $\alpha$ is related to the classical 
reorganization energy $\lambda$ by $\lambda=2\alpha\hbar\omega_c$.
A solvent with an ohmic spectral density has 
a bath polarization correlation function 
that is approximately constant at short
times and (at low temperatures)
decays algebraically at long times.  This behavior is qualitatively
similar to what has been observed in experiments for several 
solvents \cite{maroncelli} and in 
molecular dynamics simulations 
\cite{creighton88,marchi93,schulten91}.
A bath with an ohmic spectral density describes a polarization
that by itself is overdamped 
and consequently does not intrinsically exhibit vibrational coherence.  
This ensures that oscillations in the electronic populations 
whenever detected must be due solely to 
electronic coherence, but not due to
vibrational coherence of the nuclear coordinates.

It is noteworthy that the
same ohmic-like spectral density has been found by simulations 
to characterize the primary ET in bacterial photosynthesis \cite{marchi93}.  
Therefore, regarding the primary charge separation ET process, 
vibrational coherence is expected also to have no relevance.

\begin{figure}
\setlength{\unitlength}{1cm}
\begin{center}
\leavevmode
\epsfxsize 9cm 
\epsffile{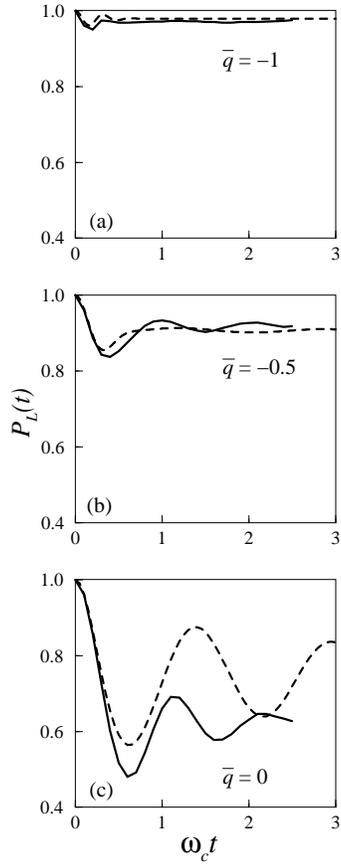}
\end{center}
\caption[]{\label{fig3}
Comparison of the data from Fig.~\ref{fig2} (dashed lines) with the 
corresponding QMC results (solid lines).} 
\end{figure}

We shall now compare QMC data to the predictions from Sect.~\ref{secii} to 
verify that the general conclusions reached there remain correct when a 
fully dynamical and dissipative bath is included in the model.
First, we examine QMC results for a symmetric ($\epsilon=0$) ET system 
with standard and shifted bath preparations.  All frequency and energy 
parameters can be expressed in dimensionless units of $\omega_c$, and
for this calculation, we have selected $\Delta/\omega_c = 4$, $\beta \hbar
\omega_c = 3$ and a friction constant $\alpha$ such that the classical
reorganization energy is $\lambda/\hbar\omega_c = 20$.  
The parameters are representative of many mixed-valence compounds, 
which generally have $\omega_c$ of the order of a hundred to several hundred 
cm$^{-1}$.  The QMC results are shown in Fig.~\ref{fig3}. 
 For a typical $\omega_c$ 
= 100~cm$^{-1}$, one $\omega_c t$ on the time-axis corresponds to 53~fs 
in real time.

The solid line in Fig.~\ref{fig3}(a) 
shows QMC data for the standard bath preparation 
$\bar q=-1$.  As expected, because the temperature and the reorganization
energy chosen put the system in the activated region, the ET rate is 
slow and the time-dependent occupation probability $P_L(t)$ decays so
slowly that on the timescale plotted it appears to remain almost constant
except for a fast initial transient.  The qualitative prediction made
in Sect.~\ref{sec:tp}
 for the same set of parameters is also shown in the same figure
as the dashed line.  Obviously, the adiabatic theory prediction captures
most of the qualitative features of the early-time dynamics of the ET.

Figure~\ref{fig3}(c) shows QMC results for a nonstandard bath preparation
with
$\bar q=0$.  This corresponds to a bath initially equilibrated to a 
zero-dipole-moment electronic state, such as the experiment depicted in
Fig.~1(b).  In this case, there are large-amplitude oscillations in the 
electronic population, indicating clearly the presence of electronic
coherence.  The close resemblance of the exact QMC results to the 
qualitative adiabatic theory prediction for the same parameters 
(dashed line) indicates that the
 reasoning given in Sect.~\ref{sec:ntp} for the 
detectability of electronic coherence in experiments with nonstandard 
bath preparations is indeed correct.
Figure~\ref{fig3}(b) shows results for an intermediate value of $\bar
q=-0.5$.
Coherent oscillations are still present, though to a smaller extent.

The discrepancies between QMC data and the adiabatic theory 
predictions visible in Fig.~3 are as expected.  
Since the adiabatic limit does not have 
dissipation, the electronic coherence persists indefinitely there,
whereas the oscillations observed in the QMC data disappear rather quickly. 
Consequently, an exact QMC calculation is necessary to estimate whether 
the coherence predicted by qualitative adiabatic theory is indeed 
preserved for a long enough period of time so that oscillations can be 
detected in experiments.
Another discrepancy between QMC and adiabatic prediction
is that the oscillation frequency in the QMC results
is initially similar to the adiabatic prediction, but then increases 
with time. This behavior is also easily rationalized 
with the picture of a dynamical polarization field which moves
away from the barrier region with time into regions where 
the difference between the energy eigenvalues is larger, 
resulting in higher oscillation frequencies.

We have shown results for only one set of parameters.
A large number of QMC calculations on other parameter sets have also 
been performed, and we found that similar coherence effects can be 
observed for many other systems as well, as long as 
$\Delta$ is similar to or larger than $\omega_c$.
Our goal is to highlight the possible electronic coherence effects that
could be detected in experiments, but not to give a detailed range of 
parameters within which electronic coherence can be observed 
(such a task would have been overwhelming).  Therefore, we will not 
explicitly show results from the other calculations here.

\begin{figure}
\setlength{\unitlength}{1cm}
\begin{center}
\leavevmode
\epsfxsize 7cm 
\epsffile{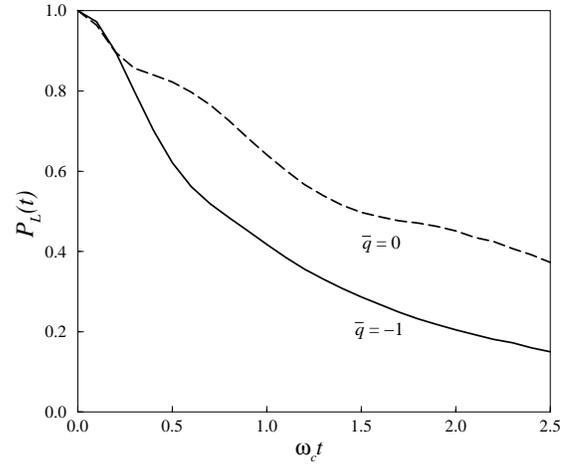}
\end{center}
\caption[]{\label{fig4}
QMC results for $P_{L}(t)$ in the activationless regime 
$\hbar\epsilon = \lambda$. Other parameters are the same as in Fig.~\ref{fig2}.}
\end{figure}

Results presented so far apply to symmetric systems.
The qualitative behavior can change quite dramatically depending on
the bias  $\epsilon$.  Figure~\ref{fig4}
 shows effects of standard and nonstandard bath
preparation in the activationless region where $\hbar\epsilon = \lambda$.
In the activationless case, 
if the bath was completely static, even a standard preparation would place
the bath inside a region where the two energy eigenstates are close to
each other, and adiabatic theory would therefore predict large coherent 
oscillations.  Clearly the QMC data in Fig.~\ref{fig4} 
for $\bar q=-1$ show no such oscillations.  
Thus, adiabatic theory delivers a qualitatively incorrect picture for this 
situation.  The reason why adiabatic theory is wrong
is not difficult to understand.  When the bath is placed right at the 
activationless crossing point on the Marcus parabola where there is no 
thermal barrier to cross from the donor to the acceptor state,
the bath should move rapidly to form the product.  
The adiabatic assumption that the bath is slow 
in the activationless region (and regions close to it) is incorrect, 
rendering the qualitative prediction of adiabatic theory invalid.
In reality, the bath being dynamical moves quickly away from the crossing 
region into regions where the difference between the energy eigenvalues is
much larger, thus destroying any sign of  coherence.

\begin{figure}
\setlength{\unitlength}{1cm}
\begin{center}
\leavevmode
\epsfxsize 7cm 
\epsffile{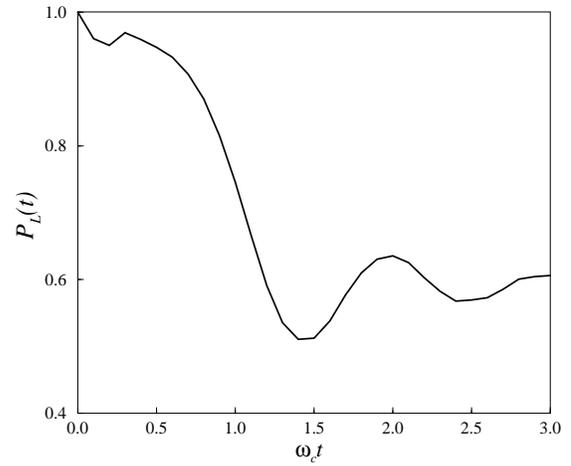}
\end{center}
\caption[]{\label{fig5}
QMC data for $P_{L}(t)$ in the case of a b-ET reaction.
Parameters  are the same as in Fig.~\ref{fig2}.}
\end{figure}

Next, we consider QMC results for the b-ET experiment depicted in
Fig.~\ref{fig1}(c). The physical situation is the following. 
For all $t<0$, the
electron is held fixed in the $|R\rangle$ state. Then, at $t = 0$, the
electron is excited to the $|L\rangle$ state by an ultrafast laser pulse,  
such that the position of the environment remains essentially unchanged, 
i.e. with the nonstandard preparation characterized by $\bar{q}=+1$.  
Figure~\ref{fig5} shows results for a b-ET experiment with the
same parameters as in Fig.~\ref{fig3}, 
except a nonstandard bath preparation of
$\bar q=+1$ was used.  Once again, coherent electronic oscillations
are clearly observable.  Although the electronic coherence is evident
in Fig.~\ref{fig5}, its origin is more subtle than the results exhibited
in
Fig.~\ref{fig3} and this deserves a closer examination.

When the system is excited to the $|L\rangle$ state, 
the bath being dynamical will move down the 
Marcus parabola toward the crossing region.
For this reason, the predictions of adiabatic theory, which assumes the
bath is completely static, are invalid just as in the
activationless case.  To understand the origin of the coherence, a
dynamical picture is required.  We focus on the sequence of events
following the vertical excitation of the system into the $|L\rangle$ state.
As the bath moves toward the crossing region, the electron being coupled
to the bath will also evolve.  There are two different scenarios.
If the bath moves slowly, the electron will have time to adjust to
the bath's position and quickly attain a mixture of the CT state
and the ground electronic state to approach one of the two energy
eigenstates.  In this case, the system would exhibit no electronic
coherence, because the electron is already in an energy eigenstate.
On the other hand, if the bath moves rapidly, the electron may 
not have
time to adjust and would stay in the $|L\rangle$ state, which
is not an energy eigenstate.
After a brief time delay, which is evident in the QMC data in
 Fig.~\ref{fig5},  
the bath polarization arrives at the 
crossing region where the two electronic energy
eigenstate are closest in energy, and the electronic population would then
exhibit large-amplitude coherent oscillations.  In other words, whether
electronic coherent oscillations are detectable in the electronic occupation
probability depends crucially on the electronic coupling $\Delta$ and the
bath dynamics timescale $1/\omega_c$.  We are in the best position to
observe electronic coherence in b-ET experiments if $\Delta$ is larger than
or comparable to $\omega_c$.

\subsection{The Equivalent Electric Field}
\label{sec:eqf}

Within the spin-boson model, the path integral formulation easily reveals
that the effects of the shifted initial preparations on the ET equal
those of an appropriately chosen time-dependent external field.
The reasoning goes as follows: When considering an initial preparation as
in EQ.~(\ref{eq:rhoq}), the time evolution of the density matrix to a time
$t$
is just a special case of the expressions given in \cite{report,weiss}, in
that the electronic variable is kept fixed at value $\bar{q}$ throughout
the imaginary time path. This leads to an influence functional, in which
the shift of the preparation only appears in a term 
$i (1+\bar{q}) \int_{0}^{t} d \tau R(\tau)\dot{\chi}(\tau)$. Here, $\chi =
q-q'$ and $q$, $q'$ are the electronic variables on the forward and
backward paths
respectively. $R$ is the imaginary part of a function $Q(z)$, which
is $(2 \mu/\hbar)^{2}$ times double integral of the correlation function
Eq.~(\ref{eq:EE}) with $Q(0)=0$.

In particular,
\begin{eqnarray}
Q(z) & \equiv &   \label{eq:Q}
 \frac{1}{\pi \hbar} \int_{0}^{\infty} d \omega
           \frac{J(\omega)}{\omega^{2}}
\\ &\times& \frac{\cosh(\hbar \beta \omega/2)
           -\cosh(\omega[(\hbar \beta/2) - iz])}{\sinh(\hbar \beta \omega/2)}
           \;. \nonumber  
\end{eqnarray}
Integrating by parts and noting that $\chi(0) = \chi(t) = 0$ 
[the electron is assumed to be in the $|L\rangle$ state initially, and
the population in state $|L\rangle$ is measured at time $t$], 
this term becomes
$-i(1+\bar{q}) \int_{0}^{\tau} d \tau \dot{R}(\tau) \chi(\tau)$. For a 
strict ohmic spectral density ($\omega_{c} \rightarrow \infty$), 
which is a good approximation for dissipative environments most often 
encountered in solid-state systems, the integral vanishes and not
surprisingly the preparation has no effect. 
This is why this kind of term   
is usually tacitly omitted in connection with
NIBA calculations \cite{leggett87,weiss}. We will see, 
however, that for parameters that are typical for many chemical systems, this 
term does have a significant effect on the ET, even in regimes where the 
NIBA is still valid, see Sect.~\ref{seciv}.
 From the Marcus picture it is clear that the prefactor, 
$1+\bar{q}$, is proportional to the displacement of 
the polarization from the standard preparation that we defined in 
Section~\ref{sec:tp}. If $\bar{q}=-1$,  
the bath is equilibrated to the donor state $|L\rangle$, i.e.~the
bath is prepared in the standard fashion. 
In this situation the term above vanishes exactly.
It is very advantageous for computations that the influence functional,
although reflecting certain initial preparations, does not include 
integrations over imaginary-time paths.

For a physical understanding, it is helpful to note that a preparation 
with $\bar{q}$ gives each path a weight 
$\exp[i(1+\bar{q}) \int_{0}^{t} d \tau \dot{R}(\tau) \chi(\tau)]$. On 
the other hand, the action of the bare TLS includes a factor
$\exp[i \int_{0}^{t} d \tau \, \epsilon \chi(\tau)]$. 
Obviously, $(1+\bar{q}) \dot{R}(\tau)$ acts just
like a time-dependent bias
that could be due to a strong electric field coupling to the dipole
moment associated with the electronic coordinate. Studying the ET  
with a nonstandard initial preparation is therefore manifestly equivalent 
to studying the ET of the same system with a standard initial bath preparation
($\bar{q}=-1$), but {\em with a time-dependent external field that mimics 
the preparation effects}. We call this field the 
``equivalent electric field''.
The physical reason for this observation is that for the ET the source of 
the field coupling to the electronic dipole moment 
is unimportant, be it due to the polarization of the 
solvent or to the external field or both.

The specific form of the equivalent external field can be seen  
by noting that the phase associated with each electronic path 
is unchanged under the transformation
\begin{equation}
(\epsilon, \bar{q}) \rightarrow 
(\epsilon+(1+\bar{q})\dot{R}(\tau),-1)\;. 
\end{equation}
Therefore, the  equivalent external field appears as an additional
term to the Hamiltonian,
\begin{equation} \label{eq:dh}
\Delta H =  \frac{\hbar}{2}(1+\bar{q}) \dot{R}(t) \;
    \biggl(|L\rangle \langle L| - |R \rangle \langle R| \biggr)\;. 
\end{equation}
Eq.~(\ref{eq:dh}) establishes the connection between nonstandard initial 
preparations and the work already done on driven spin-boson systems,
see Refs.\cite{milena,dak,dak-co,coalson-prl}.
It is interesting to note that Eq.~(\ref{eq:dh})
decreases in time just like the classical
damping kernel $\gamma(t)$ \cite{weiss}, since from Eq.~(\ref{eq:Q}) 
\begin{equation} \label{damp}
\gamma(t) = (2/\pi) \int_{0}^{\infty} d \omega
  \frac{J(\omega)}{\omega} \cos(\omega t) = 2 \hbar \dot{R}(t) \;.
\end{equation}

\begin{figure}
\setlength{\unitlength}{1cm}
\begin{center}
\leavevmode
\epsfxsize 7cm 
\epsffile{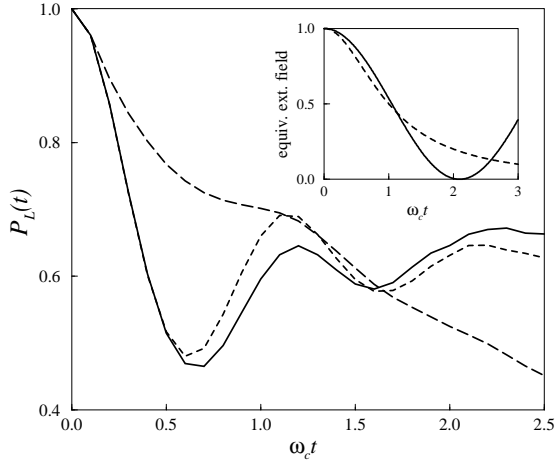}
\end{center}
\caption[]{\label{fig6}
$P_{L}$ for the standard initial preparation but with an equivalent 
external field. In order to reproduce ET of the $\bar{q}=0$ initial 
preparation (short-dashed curve given here for comparison), 
the external field must take the form of 
the dashed line in the inset. If the equivalent field is approximated 
by a sinusoidal field on a static background 
(solid line in inset), the ET (solid line) behaves very
similarly to the nonstandard preparation showing large electronic coherence.
The long-dashed line is
an estimate of the population if a random-phase
cosine-modulated external field is used instead (see text).} 
\end{figure}

So far, the equivalent external field has just been a name for 
the effect that nonstandard distributions have on the influence 
functional. However, as mentioned in Sect.~\ref{sec:ntp}, the 
equivalent external field may actually be applied to the system 
to mimic a nonstandard initial preparation in an experiment in which the 
actual preparation is the standard one, as in Figs.~\ref{fig1}(d) and (e).
To give a flavor, we want to repeat the
calculation in Fig.~\ref{fig3} for
the standard preparation but under the influence of an equivalent external 
field. The proper 
form of the field is given by Eqs.~(\ref{eq:Q}) and (\ref{eq:dh}) and 
shown in the inset of Fig.~\ref{fig6}
as the dashed line. In a real experiment, it is much easier to
generate a sinusoidal field.  Therefore, in the QMC calculation, we 
used a simple cosine field plus a static bias shown as the solid line 
in the inset of Fig.~\ref{fig6}.  With this approximate equivalent
electric
field, we see from the solid line in Fig.~\ref{fig6} 
that the external field induces a behavior very
similar to the $\bar{q} = 0$ initial preparation.

Instead of applying the cosine external field one might try 
something even simpler and couple a CW source into the system where the 
phase at $t=0$ is selected at random. The
 long-dashed curve in Fig.~\ref{fig6}
depicts
the average of the electronic population from four different calculations,
in which the 
external fields all had a cosine time-dependent modulation but
had phases shifted by 0, 1/4, 1/2 and 3/4 of a period. Although much 
weaker, coherent oscillations are still clearly evident.  A
typical field strength required would be about $10^{4} KV/cm$, which is not
difficult to achieve.

What we have shown up to this point is that given specific
initial preparations, electronic coherence can lead to oscillations in the
electronic populations that are pronounced enough to be detected in
experiments.  However, a femtosecond time-resolved spectroscopic analysis 
of the electronic populations is not a trivial experiment and the 
interpretation of the signal is often complicated by 
effects unrelated to the ET itself.
Moreover, in some systems where we would expect
the ET to be coherent, the oscillations are on the timescale of tens of  
fs, and therefore cannot be resolved even in the fastest pump-probe
experiments possible today.  Finally, we want to point out that our
calculations are based on the assumption of a
gaussian bath, neglecting e.g.~the role of strong anharmonic modes that
couple to the ET.  Clearly, nature is more complicated than our model, but
it is safe to conclude that strong dissipation by itself does not prevent
electronic coherence from being observed.  In spite of the above limitations,  
we hope that our results will  
encourrage experimental attempts to detect electronic coherence in the
scenarios we have suggested.

\section{Initial Preparation Effects for Nonadiabatic Electron Transfer}
\label{seciv}

In this section, we briefly discuss the effect of nonstandard 
initial preparations in the nonadiabatic limit,
$\Delta \ll \omega_{c}$. The effects should be smaller in this limit, because 
as argued above when the timescale for a electron hop, $1/\Delta$, 
is much longer than the typical timescale of 
of the bath, $1/\omega_{c}$, the initial bath state 
is not expected to matter.  Studying the effects of shifted
initial preparations in the nonadiabatic limit becomes especially
interesting in the context of vibrational coherence, because a 
polarization distribution shifted away from the energy surface minimum may
lead to an interesting dynamical behavior.
 
In the nonadiabatic limit, the ET transfer can be very accurately 
described in the framework of the non-interacting blip approximation (NIBA) 
\cite{leggett87}. Using the concept of the equivalent electric field, 
a nonstandard preparation can be easily 
incorporated in the NIBA. In fact, the proper form of the NIBA expression 
for $\dot{P}_L(t)$ in the 
presence of a driving field has already been given, e.g.,
in Refs.\cite{milena,coalson-prl}, 
though from a very different perspective.  All we need 
to do is to modify these expressions by choosing the 
interaction energy $\Delta H(t)$ according to the external field 
in the form (\ref{eq:dh}),  getting
\begin{eqnarray}
\lefteqn{\dot{P}_L(t)=-(\Delta^{2}/2) \int_{0}^{t} d \tau e^{-S(\tau)}  
        \biggl[\sin(R(\tau)) \times \biggr.}  \nonumber \\
 & &    \sin \biggl(\epsilon \tau
        + (1+ \bar{q})[R(t) - R(t- \tau)]\biggr)+ \cos(R(\tau)) 
        \times  \nonumber \\
 & &    \biggl. \cos \biggl(\epsilon \tau +( 1+
       \bar{q})[R(t) - R(t-\tau)]\biggr) (2 P_L(t-\tau)-1)\biggr ] \;.
        \label{eq:xdot}
\end{eqnarray}

We first examine the short time behavior of the ET. For times
smaller than $1/\omega_{c}$, $R(\tau)$ is proportional to the
reorganization energy, $\lambda$,
and at high temperatures, $S(\tau) = \lambda\tau^2/\hbar^{2} \beta$.
With $P_L(t) \approx 1$ for $t \ll 1/\omega_{c}$, one finds
\begin{equation} 
\Gamma(t) \equiv -\frac{\dot{P}_L(t)}{P_L(t)} \approx  \int_{0}^{t} d\tau 
e^{-\lambda \tau^2/\hbar^2 \beta} \cos ([(\lambda/\hbar) \bar{q}
+ \epsilon] \tau)\;. 
\end{equation}
The ``rate'' $\Gamma(t)$ defined here displays a
few oscillations of frequency $[(\lambda/\hbar)\bar{q} +\epsilon]$
at extremely short times.
This frequency corresponds to the vertical distance of the diabatic Marcus 
parabolas at $\mu {\cal E}=\bar{q} \lambda/2$, so 
these oscillations are due to electronic resonance. 
In the nonadiabatic 
limit, the ET takes place on a much longer timescale compared 
to this resonance.  As a result, although these
oscillations can be observed in $P_L(t)$, they exist only for very
short times (up to $\approx 1/\omega_c$) and their amplitudes are only 
$1 \%$ of $P_L(t)$.  Therefore, electronic coherence plays only a minor role
in nonadiabatic ET.

The more interesting question is whether a nonstandard initial
bath distribution causes effects on timescales $t \approx 1/\omega_{c}$. 
Experiments in the photosynthetic reaction center
have given evidence of vibrational coherence on the same timescale as
the ET itself \cite{vos}, 
where oscillations of the nuclear coordinate distribution (a ``wavepacket'')
within the donor energy parabola could be observed before
dephasing occurred. Since the nuclear coordinate ${\cal E}$ must lie
in the Landau-Zener crossing region for ET to occur, the nuclear motions
should result in oscillatory transfer rates and in the electronic occupation 
probabilities.  As already pointed out,
with an ohmic bath there is no vibrational coherence.
However, one should still be able to detect the relaxation of the polarization
from the transfer rate.
To test this wavepacket picture, 
we chose to investigate a rather small electronic coupling,
$\Delta = 0.4 \omega_{c}$, with $\lambda = 3 \hbar \omega_{c}$ and 
$\hbar \beta \omega_{c} = 10$.

\begin{figure}
\setlength{\unitlength}{1cm}
\begin{center}
\leavevmode
\epsfxsize 7cm 
\epsffile{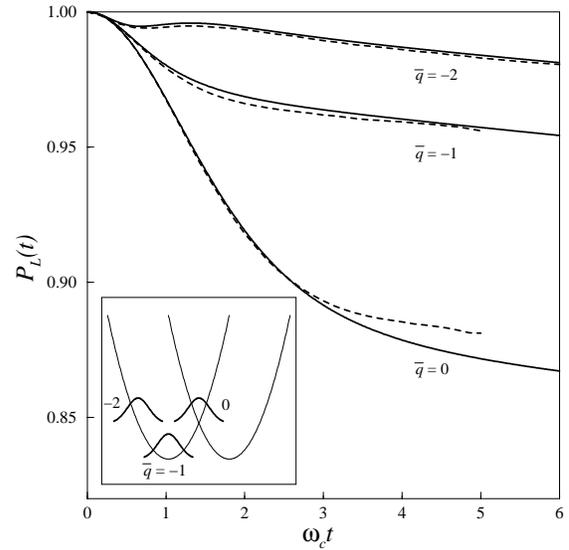}
\end{center}
\caption[]{\label{fig7}
$P_{L}(t)$ for $\Delta = 0.4 \omega_{c}$, $\lambda = 3 
\hbar \omega_{c}$ and $\hbar \beta \omega_{c} = 10$
for initial bath preparations 
with $\bar{q} = 0,-1,-2$ in the symmetric case $\epsilon = 0$. The 
solid curves have been obtained in a NIBA calculation, the short-dashed
curves by QMC. The inset shows the position of the corresponding initial 
polarization distributions in the Marcus picture.}
\end{figure}

\begin{figure}
\setlength{\unitlength}{1cm}
\begin{center}
\leavevmode
\epsfxsize 7cm 
\epsffile{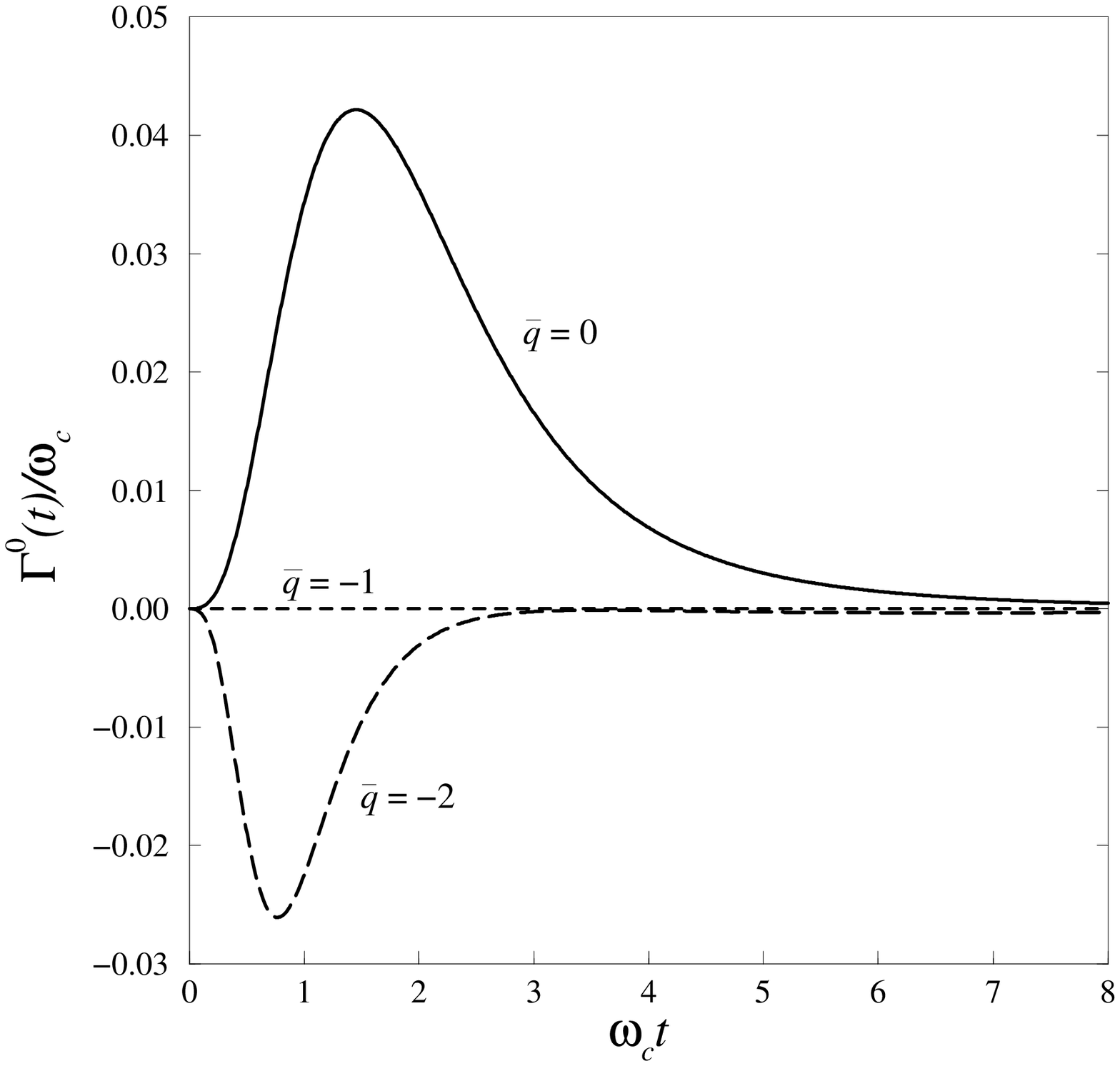}
\end{center}
\caption[]{\label{fig8} The normalized transfer rates $\Gamma^0(t)$
for the situation of Fig.~\ref{fig7}.} 
\end{figure}

Figure~\ref{fig7} shows $P_L(t)$ for 
$\bar{q} =0,-1$ and $-2$ obtained from a NIBA 
calculation (solid lines), and from
quantum dynamics simulations (dashed lines). 
$\bar{q}=-1$ corresponds to standard prepartion with the bath 
equilibrated to the donor state. $\bar{q} = 0$ 
corresponds to a polarization distribution 
initially placed at the crossing point of the Marcus parabolas (inset of 
Fig.~\ref{fig7}) where the majority of the ET occurs.
The ET proceeds rapidly. Finally, $\bar{q}=-2$ 
corresponds to a polarization distribution centered
on the opposite side of the minimum. To get 
to the crossing region, the bath has to first relax to the 
parabola minimum, causing a delay in the ET. 
This is most easily discerned from the rate $\Gamma(t)$.
Figure~\ref{fig8} shows the difference of the rates for 
$\bar{q}=-2$ and 0 from the standard preparation $\bar{q}=-1$.
They are clearly out of phase with each other, in agreement with
the picture of two wavepackets starting at opposite turning points. 

\begin{figure}
\setlength{\unitlength}{1cm}
\begin{center}
\leavevmode
\epsfxsize 7cm 
\epsffile{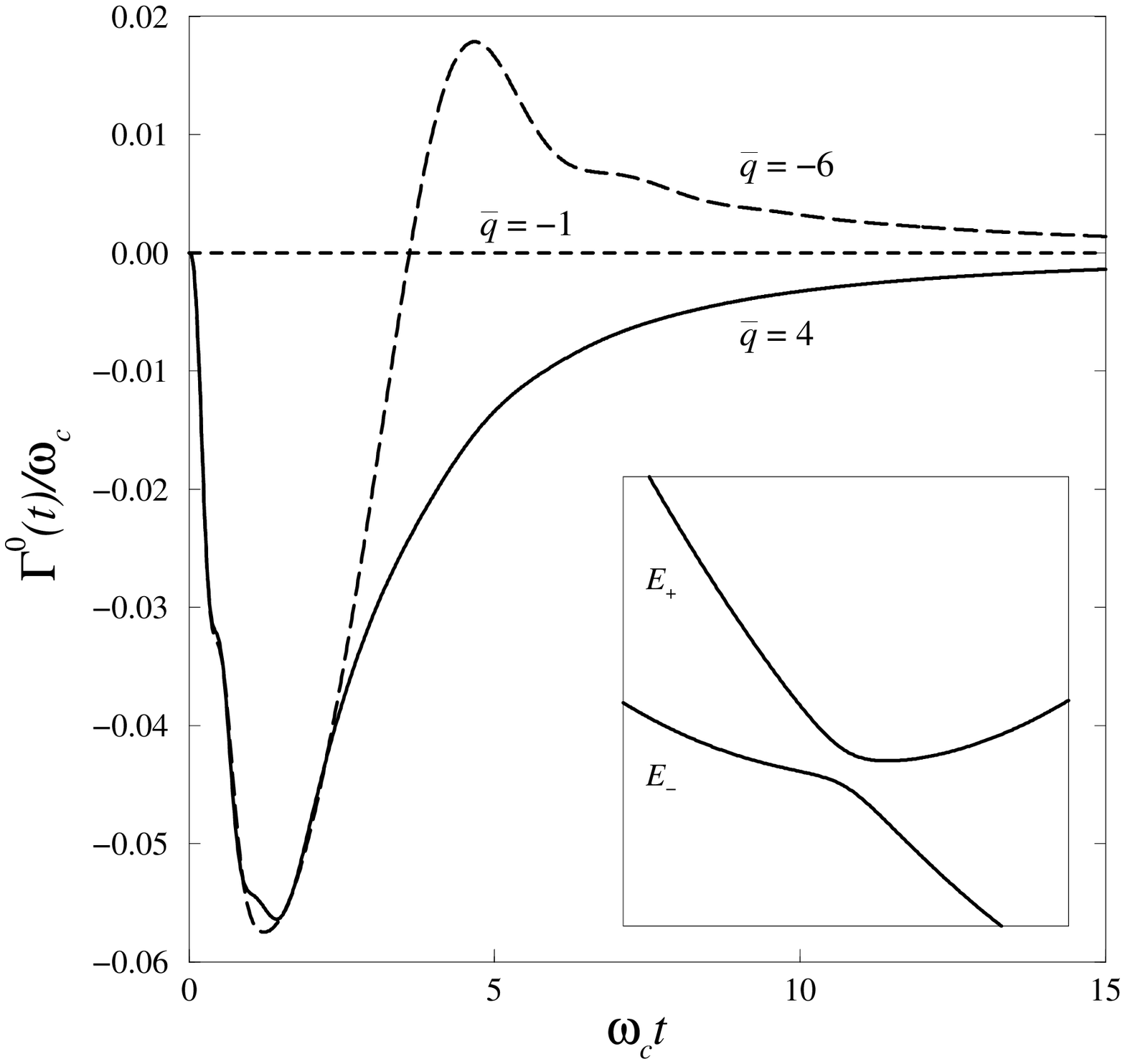}
\end{center}
\caption[]{\label{fig9} The NIBA-calculated
normalized transfer rates in the
activationless regime $\hbar\epsilon=\lambda$. The 
wavepackets start at opposite sides of the donator 
parabola minimum. The inset shows a 
closeup look of the Landau-Zener crossing region, which allows
to explain the asymmetric 
behavior in the wavepacket picture.} 
\end{figure}

A different situation is found in the activationless regime, where 
the Landau-Zener crossing region is now at the bottom of the donor parabola. 
Two wavepackets
starting at opposite turning points should result in 
rates that are in phase. Figure~\ref{fig9} shows that this is indeed the 
case for times up to $1/\omega_{c}$. Once the polarization
reaches the crossing region, 
the two situations are no longer symmetric. The inset of 
Fig.~\ref{fig9} shows a closeup of the crossing region.
Compared to the one approaching from the right ($\bar{q}=4$), 
the polarization coming from the left side $(\bar{q} = -6)$ will have 
an easier time tunneling, by simply
sliding down the lower energy surface.  This is reflected by the 
transfer rates shown in Fig.~\ref{fig9}.
We see that in the nonadiabatic regime nonstandard 
preparations also have a visible effect on the ET 
consistent with a simple wavepacket picture.
However, the effect is much less significant than in the adiabatic regime.

Whether vibrational coherence causes oscillations in the populations or not
depends on the specific spectral density that is involved. If the
spectral density is dominated by a narrow range of frequencies, not 
surprisingly we find oscillations in $P_L(t)$ when a nonstandard
initial preparation is used. These oscillations
vanish when the polarization distribution
is initially placed at the minimum of the donor parabola.
The effect of vibrational coherence on the ET can be predicted 
by looking at the relaxation behavior of the initially shifted
polarization distribution, i.e. the form of 
$\dot{R}(t)$ or the damping kernel. For the
photosynthetic reaction center,
the classical correlation functions have been determined by molecular
dynamics
simulations \cite{marchi93}. Although they show some high frequency 
oscillatory modulations, the basic behavior is an overdamped relaxation.
We conclude that vibrational coherence should {\em not} cause
significant 
oscillations in the electron populations in the reaction center, even if the 
initial wavepacket is largely displaced. This is also the conclusion from 
quantum dynamics simulations employing the spectral density determined in 
\cite{marchi93}, and in agreement with recent experimental
findings
\cite{zinth}, while another
group suggests the opposite \cite{streltsov}. 

Not surprisingly, taking a model spectral density containing
a large concentration of weights within a very narrow range of frequencies,
our NIBA and QMC calculations both showed that $P_L(t)$ becomes
oscillatory.

\section{Concluding Remarks}

The unambiguous detectability of electronic coherence in 
ET reactions is still an open question
from an experimental point of view.  In this paper,
we have shown that there are situations, in
which one might be able to indeed observe oscillatory behaviors
in electronic occupation probabilities which are caused
by electronic coherence.  The most appropriate systems
seem to be symmetric mixed-valence compounds, which are close to
the adiabatic ET regime.  The crucial
point is then to consider preparations where 
the solvent polarization initially forms a distribution centered 
near the Landau-Zener crossing region.  We have given some examples
for how one can create such an initial preparation in practice.
Before the wavepacket slides away, i.e. before dephasing occurs, the
electron can have enough time to oscillate back and forth
several times between the electronic surfaces.  The resulting large-amplitude
oscillations should be easily detectable in experiments. 

Besides the experimental relevance, we believe that these
results might be of use for other theoretical studies of nonequilibrium
preparations in condensed phase systems.
We have extended the Feynman-Vernon influence
functional method to account for nonstandard initial preparations.

From the path-integral expressions, one can easily deduce that such
nonstandard preparations can be thought of as an equivalent
external field. This establishes a link with the field of
driven dissipative quantum systems and this connection is useful in actual
computations.  Another important consequence of this concept
is that one can quite easily decide whether a particular 
ET system may exhibit oscillatory rates by inspecting the
equivalent external field.  If the latter does not show 
oscillations but simply relaxes to zero, the polarization
motion will be overdamped, and oscillatory behaviors must have
an electronic origin.  For that reason, we conclude that vibrational coherence
cannot cause oscillations in the occupation probabilities
for a large class of common ET systems. A prominent example is the
primary ET step in the photosynthetic reaction center.

To conclude, we hope that this paper will stimulate experiments
investigating electronic coherence in electron transfer reactions,
as well as theoretical studies dealing with the effects of 
special initial preparations, e.g.~due to the laser pulse
excitation of a wavepacket, in condensed phase systems.

\acknowledgments

Throughout this work, we have benefitted from many enlightening 
discussions with Stephen Bradforth and Gerhard
Stock.
This research has been supported by the National Science Foundation
under grant CHE-9528121 and by 
 the Schwerpunkt ``Zeitabh\"angige
Ph\"anomene und Methoden in Quantensystemen in der Physik und 
Chemie'' of the Deutsche Forschungsgemeinschaft (DFG).
CHM is a NSF Young Investigator (CHE-9257094), 
a Camille and Henry Dreyfus Foundation Camille Teacher-Scholar and a 
Alfred P. Sloan Foundation Fellow.  
CHM acknowledges support from the DFG Schwerpunkt 276 during an
extended stay at Freiburg.
JA is a Feodor Lynen Fellow of the Humboldt Foundation.
Computational resources have been 
provided by the IBM Corporation under the SUR Program at USC.

\begin{thebibliography}{99}
\bibitem{remark2} We note that bath modes with frequencies
           $\omega \gg \Delta$ can actually be incorporated by
           an adiabatic elimination procedure \cite{leggett87,stu}, 
           which simply leads to a renormalization of $\Delta$. 
           In what follows, we imagine that such high-frequency 
           modes have already been eliminated.
\end{thebibliography}
\end{document}